\documentclass[aps,prl,showpacs,preprintnumbers,amsmath,amssymb,superscriptaddress,twocolumn]{revtex4-1}
\usepackage{amsmath}
\usepackage{graphicx}
\usepackage{hyperref}
\usepackage{xcolor}
\usepackage{float}
\usepackage{ulem}
\usepackage{braket}
\usepackage{comment}

\graphicspath{{./figures/}{./figures_supp/}}

\definecolor{gold}{rgb}{0.83, 0.69, 0.22}

\newcommand{\bs}{\boldsymbol}
\newcommand{\lc}{\rho_2^B(\bs{r})}

\newcommand{\ssec}[1]{\textit{#1}.---}
\newcommand{\midrule}{\hline}
\newcommand{\bottomrule}{\hline\hline}

\begin{document}

\title{Ferromagnetic semimetal and charge-density wave phases of \\
interacting electrons in a honeycomb moir\'e potential}

\author{Yubo Yang}
\affiliation{Center for Computational Quantum Physics, Flatiron Institute, New York, NY, 10010, USA}
\author{Miguel A. Morales}
\affiliation{Center for Computational Quantum Physics, Flatiron Institute, New York, NY, 10010, USA}
\author{Shiwei Zhang}
\affiliation{Center for Computational Quantum Physics, Flatiron Institute, New York, NY, 10010, USA}
\date{\today}
\begin{abstract}
The exploration of quantum phases 
in moir\'e systems 
has drawn intense
experimental and theoretical efforts. 
The realization of honeycomb symmetry
has been a recent focus.
The combination of strong interaction and honeycomb symmetry can lead to exotic electronic states such as fractional Chern insulator, unconventional superconductor, 
and quantum spin liquid.
Accurate computations in
such systems, with reliable treatment of strong long-ranged Coulomb interaction
and approaching the large system sizes 
to extract thermodynamic phases,
are mostly missing.
We study the two-dimensional electron gas on a honeycomb moir\'e lattice at quarter filling, using fixed-phase diffusion Monte Carlo.
The ground state 
phases of this important model 
are determined in the parameter regime relevant to 
current experiments. With increasing moir\'e potential,
the systems transitions from a
paramagnetic metal to 
an itinerant ferromagnetic semimetal and then a charge-density-wave insulator.
\end{abstract}
\maketitle

\ssec{Introduction}
When two monolayers of crystalline patterns are stacked with a small misalignment, their interference creates a moir\'e superlattice with long wavelength.
This allows the long-range Coulomb interaction to dominate when a dilute gas of carriers is doped into the bilayer.
Recently, transition metal dichalcogenide (TMD) devices have become highly productive quantum simulators of strongly interacting physics~\cite{kennes_moire_2021,mak_semiconductor_2022}.
Many exotic electronic phases have been realized, including
generalized Wigner crystal (GWC)~\cite{wang_correlated_2020,regan_mott_2020,shabani_deep_2021,li_imaging_2021-1,nieken_direct_2022,xiang_quantum-melting_2024},
Kondo heavy fermion liquid~\cite{guerci_chiral_2023,zhao_gate-tunable_2023},
kinetic magnetism~\cite{tao_observation_2023-1,ciorciaro_kinetic_2023},
fractional Chern insulator~\cite{cai_signatures_2023,jia_moire_2023,zeng_thermodynamic_2023}, and unconventional superconductor~\cite{xia_unconventional_2024}.

Charge carriers injected into a semiconductor heterostructure are 
well described
by a two-dimensional electron gas (2DEG)~\cite{Ando_electronic_1982}.
Recent realization of the 2DEG in thin semiconductors include 
ZnO heterostructure~\cite{falson_competing_2022}, AlAs quantum well~\cite{hossain_observation_2020,hossain_spontaneous_2021,hossain_anisotropic_2022}, and MoSe$_2$~\cite{smolenski_signatures_2021,zhou_bilayer_2021,Sung2023}.
Interlayer coupling in
a bilayer TMD device varies in the moir\'e unit cell and effectively imposes an external periodic potential on the 2DEG.
Given the triangular symmetry of the TMD monolayers, the external moir\'e potential typically shares the triangular ($C_3$) symmetry of the underlying atomic lattice.
However, in special cases, honeycomb ($C_6$) symmetry can emerge. 

The possibility of realizing 
honeycomb 
lattices has been a focus of recent studies~\cite{angeli2021,angeli_twistronics_2022,kaushal_magnetic_2022,pan_realizing_2023}.
At the moir\'e length scale, the emergent Dirac band crossings at the $K$ points can be modified by the Coulomb interaction and give rise to interesting physics.
At fractional filling of the honeycomb lattice, long-range interaction is crucial for understanding the GWC phases that arise in the strongly interacting limit.
The relative strength of long- and short-range interactions was shown to change the nature of the magnetic interaction from ferromagnetic to anti-ferromagnetic~\cite{kaushal_magnetic_2022}.
Experimental efforts have been rapidly advancing in this direction.
For example, a switchable ferromagnetic state~\cite{anderson_programming_2023} along with a  fractional Chern insulator~\cite{cai_signatures_2023} were realized in twisted bilayer MoTe$_2$ devices, while superconductivity was very recently reported in twisted bilayer WSe$_2$~\cite{xia_unconventional_2024}.

In these systems, the rich set of candidate ground states arise from delicate competition and cooperation between band structure, moir\'e potential, and the interaction.
Treating such a correlated electron system is intrinsically hard, and 
remains an outstanding general problem.
It is challenging for
theoretical tools to 
have both high accuracy and low 
computational scaling, which are often needed
to resolve the relative stability among the different candidate orders or phases 
to give reliable predictions.  
As we have seen from even the simplest examples of correlated systems such as the Hubbard model \cite{LeBlanc_Solutions_2015,
Mingpu_Absence_2020,Hao_Coexistence_2024},
accurate computational results are indispensable, for benchmarking simpler methods, 
validating and enhancing theoretical understanding, and making connections with experiments.

In this paper, we study the effect of a moir\'e potential with honeycomb symmetry on the 2DEG in the presence of strong electron-electron Coulomb interaction.
We employ the diffusion Monte Carlo (DMC) method 
to accurately treat the moir\'e continuum model.
The results from this correlated many-body method are beyond the reach of 
independent-electron approaches and 
have not been quantified or observed in this system.
We find a rich ground-state phase diagram at quarter filling, including a transition from a paramagnetic metal into a 
ferromagnetic semimetal phase, before 
a transition into a
GWC, an insulating
ferromagnetic charge-density wave state.

\ssec{Model and Methods}
The moir\'e continuum hamiltonian~\cite{Wu2018,Wu2018a} (using the Wigner Seitz radius $a$ as length unit and kinetic energy scale $W=\frac{\hbar^2}{2ma^2}$ as energy unit) 
\begin{align} \label{eq:mch-rs}
\mathcal{H} = -\frac{1}{2}\sum\limits_i \nabla_i^2 - \lambda\sum\limits_i \Lambda(\bs{r}_i) + r_s\sum\limits_{i<j} \dfrac{1}{\vert \bs{r}_i-\bs{r}_j\vert}
\end{align}
describes a 2DEG with interaction strength $r_s=a/a_B$ in an external moir\'e potential with depth $\lambda=V_M/W$.
The Bohr radius $a_B$ is set by the effective mass of the electrons $m$ and a dielectric constant.
We use the leading-order approximation to the moir\'e potential in reciprocal space  $\Lambda(\bs{r})=\sum\limits_{j=1}^3 2\cos(\bs{r}\cdot\bs{g}_j+\phi)$, where $\bs{g}_j$ are three of the smallest non-zero reciprocal lattice vectors of the moir\'e unit cell.
The shape of the moir\'e potential is controlled by a single parameter $\phi$.
Honeycomb symmetry can be obtained at $\phi=60^\circ+120^\circ m$, $m\in\mathcal{Z}$. 

We use fixed-phase diffusion Monte Carlo (FP-DMC) to find the ground-state of eq.~(\ref{eq:mch-rs}).
Details of the methodology are available in Refs.~\cite{Ortiz1993,Foulkes2001,dmc_details}.
We emphasize that FP-DMC is variational and works directly in the complete-basis-set limit.
This approach has been the
computational method of choice in the electron gas~\cite{ceperley_ground_1980,attaccalite_correlation_2002,drummond_phase_2009}, and has
demonstrated excellent accuracy in related systems~\cite{yang_mit_2024}.
The method can 
be applied 
in the continuum 
with hundreds of electrons, so we can reach large simulation cells 
when calculating properties, and draw conclusions more reliably about 
the thermodynamic limit.

\begin{figure}[!htbp]
\includegraphics[width=\linewidth]{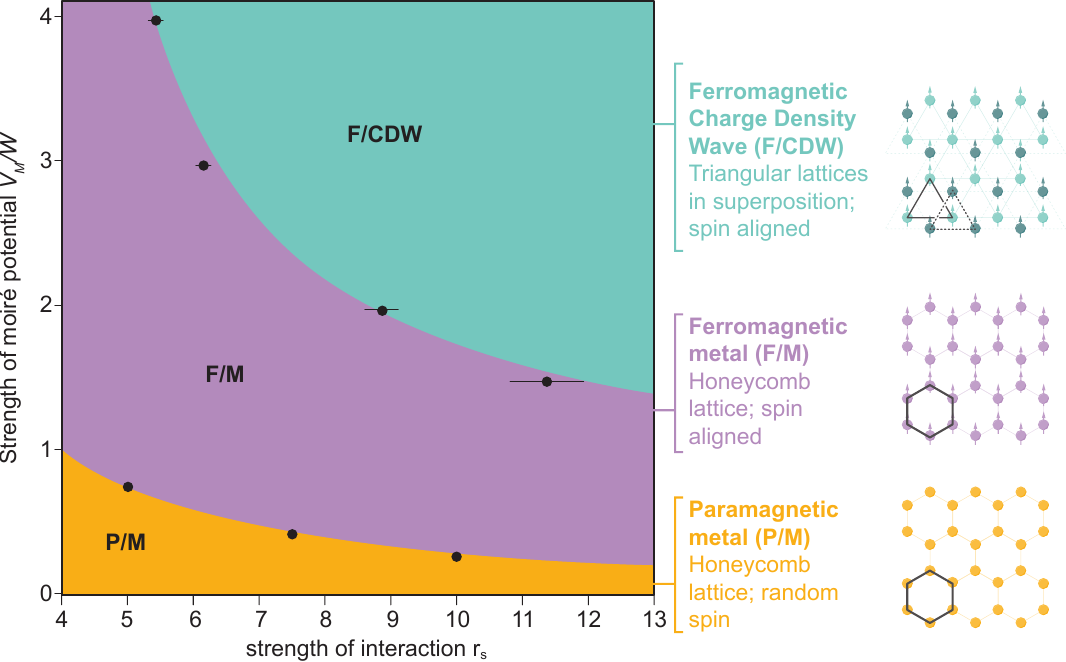}
\caption{Ground-state phase diagram of 
the honeycomb moir\'e continuum hamiltonian at quarter filling (one electron per moir\'e unit cell).
As the moir\'e potential and the electron interaction strengths increase, 
the system exhibits, progressively, three distinct phases: a paramagnetic metal (P/M), a ferromagnetic metal (F/M), and a ferromagnetic insulator
with charge density wave order (F/CDW),
as illustrated by the cartoons. 
Black symbols are 
transition points 
determined by our QMC 
calculations.
}
\label{fig:phase}
\end{figure}

\ssec{Results and Discussions}
In the absence of a moir\'e potential, even under strong Coulomb interaction (e.g., $r_s \sim 30$), the 2DEG is expected to remain a paramagnetic liquid~\cite{drummond_phase_2009}.
Therefore, in the experimentally relevant range $5 < r_s < 15$, an external potential is needed to induce long-range charge and magnetic correlations.
In the presence of a triangular moir\'e potential with commensurate wavelength, a GWC is expected to be stable~\cite{yang_mit_2024}.
However, the effect of a moir\'e potential with honeycomb symmetry is more subtle.
Figure~\ref{fig:phase} summarizes our results. With
increasing interaction and moir\'e potential depth, we find three distinct phases: a paramagetic metal (P/M), a ferromagnetic semimetal (F/M), and a ferromagnetic charge density wave (F/CDW).
Below we quantify and characterize each phase in more detail. As an overview, 
when 
the kinetic energy dominates, the P/M phase is almost identical to the unperturbed 2DEG. It is nearly isotropic with only a small density modulation induced by the moir\'e potential.
As the potential deepens, we find an abrupt transition into the F/M phase, where the spins spontaneously polarize and the Fermi surface
shrinks to
the K points of the Brillouin zone.
From the F/M phase, sufficiently strong interaction can set off a triangular charge density wave, which strongly breaks the honeycomb sublattice symmetry in pair correlations.
This insulating F/CDW phase is more isotropic than the F/M phase due to the lack of metallic directions.
Qualitatively, it can be visualized as the equal superposition of two triangular GWCs pinned to the two different sublattices of the honeycomb as shown 
by the depiction at the top right of Fig.~\ref{fig:phase}.

\begin{figure}[ht]
\includegraphics[width=\linewidth]{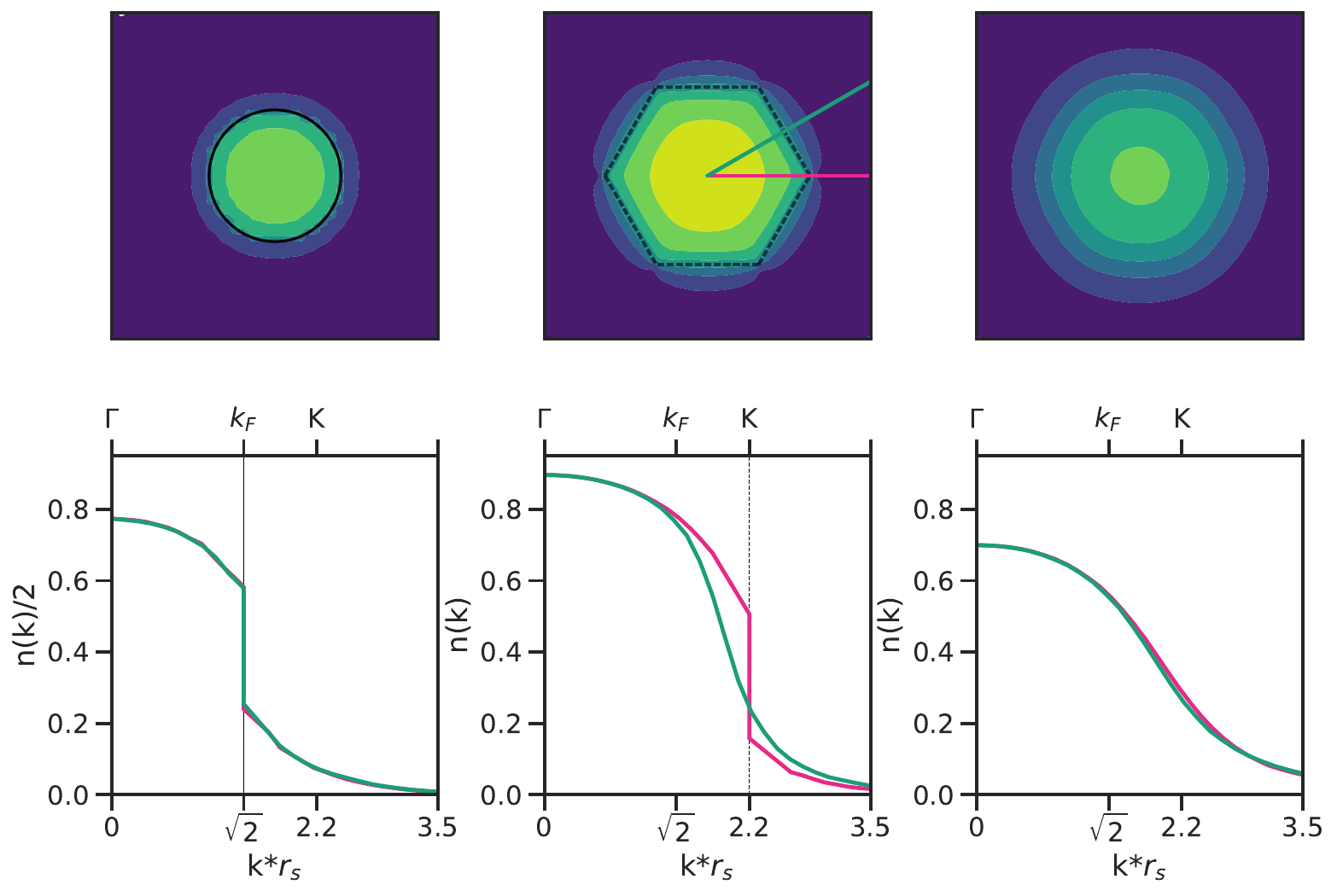}
\caption{Momentum distribution at fixed $r_s=7$ 
for three representative moir\'e potential depths:
$V_M/W=0.3$, 
$1$, and $3$,  
for the P/M,
the F/M, 
and the F/CDW,
respectively.
Top panels show 2D density maps, while the bottom panels show 1D linecuts along $\Gamma K$ (pink) and $\Gamma M$ (green).
Note the nearly isotropic discontinuity in the P/M,
the jump at only the $K$ points in the F/M,
and the smooth $n({\mathbf k})$ in the F/CDW insulator.
}
\label{fig:nofk}
\end{figure}

The three observed phases exhibit distinct signatures in momentum space.
Figure~\ref{fig:nofk} shows the momentum distribution $n(\bs{k})$ of the three phases.
In the P/M phase, $n(\bs{k})$ is only weakly perturbed by the moir\'e potential.
The nearly circular Fermi surface remains largely undistorted.
Relative to the unperturbed 2DEG (Fig.~\ref{fig:dmc-p60-nk-heg}), we find that the moir\'e potential scatters
low-momentum states within the Fermi surface towards secondary Fermi surfaces and the high-momentum tail.
In contrast, the response of the ferromagnetic (FM) semimetal to the external potential is highly anisotropic.
The Fermi surface is destroyed everywhere except for at the $K$ and $K'$ points.
No secondary Fermi surface is observable along the $\Gamma K$ and $\Gamma M$ directions.
Compared to the momentum distribution of a polarized Fermi liquid, a substantial and highly anisotropic shift of momentum density can be observed, transitioning from just below $k_F$ to just above $k_F$.
Along $\Gamma M$, the redistribution follows a pattern typical of a metal-to-insulator transition, whereas along $\Gamma K$, it is more akin to that of a paramagnetic to ferromagnetic metal-to-metal transition.
Upon reaching the insulating phase, the Fermi surfaces of the semimetal disappear and
$n(\bs{k})$ becomes a smooth function.

\begin{figure*}[!htbp]
\includegraphics[width=\linewidth]{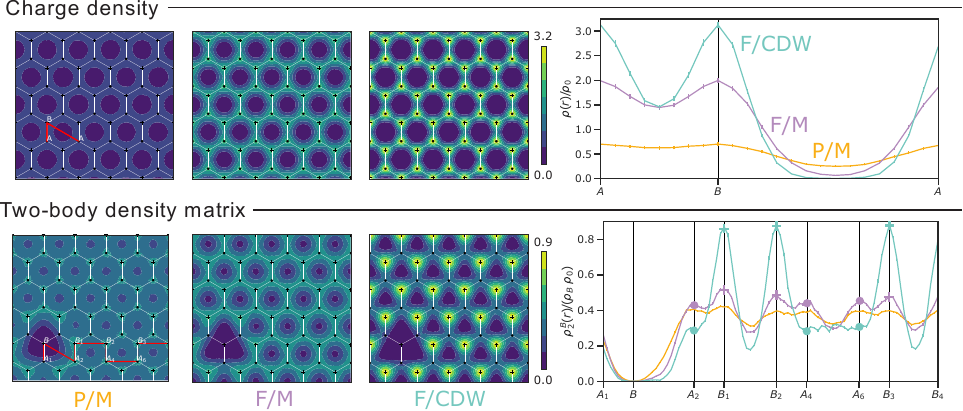}
\caption{Charge density (top panels) and 
the pair distribution function $\lc$ of eq.~(\ref{eq:rho2a}) (bottom panels) of the three phases at the same conditions as Fig.~\ref{fig:nofk}.
The charge densities of all three phases show equal occupation of the honeycomb A (black dots) and B (black pluses) sublattices.
As quantified by the line cuts, deeper moir\'e potentials induce stronger density modulations.
$\lc$ exhibits 
an xc hole 
around $\bs{r}=\bs{0}$. 
Sublattice symmetry is restored away 
from the hole in the P/M and F/M phases.
In the F/CDW phase, strong sublattice symmetry breaking is stable at large separations.
To quantify the this, 
$\lc$ is plotted along
the line cut (red line in the bottom left panel) 
which goes through four pairs of sublattice sites at increasing distances away from the hole.
In the P/M phase, no indication of sublattice symmetry breaking can be found away from the xc hole.
In the F/M phase, a small amount of sublattice symmetry breaking, which is evident at the nearest-neighbor $B_1$ site, decays with distance.
In the F/CDW phase, strong sublattice symmetry breaking can be observed at all distances.
}
\label{fig:dens-pair}
\end{figure*}

The metal-insulator transition within the fully polarized sector is subtle.
As shown in the top panels of Fig.~\ref{fig:dens-pair}, direct inspection of the one-body density 
\begin{equation} \label{eq:rho}
\rho(\bs{r}) = N\int d\bs{r}_2\dots d\bs{r}_N \vert \Psi_N(\bs{r},\dots,\bs{r}_N) \vert^2
\end{equation}
is not sufficient for identifying the metal-insulator transition from F/M to F/CDW.
In an independent-electron solution, for example from Hartree-Fock or 
a density-functional theory calculation, a CDW phase would 
break  symmetry in the charge density.
In the many-body solution, however,
the ground-state charge density retains sublattice symmetry on the honeycomb sites 
in both phases, despite gap openings at the Dirac points in the insulating F/CDW phase.
To further characterize and 
quantify them, 
we analyze the two-body correlation
\begin{equation} \label{eq:rho2}
\rho_2(\bs{r}_1,\bs{r}_2) = N(N-1) \int d\bs{r}_3\dots\bs{r}_N \vert \Psi_N(\bs{r}_1,\dots,\bs{r}_N) \vert^2.
\end{equation}
By fixing one of the positions in $\rho_2$ to the vicinity $\mathcal{V}$ of a particular 
lattice site, we can examine the charge correlation function 
with respect to this site.
With no loss of generality (due to translational symmetry), we choose a site on the ``B'' sublattice, 
$\bs{R}_B$, and compute:
\begin{equation} \label{eq:rho2a}
\lc = \int_{\bs{r}_1\in\mathcal{V}(\bs{R}_B)} d\bs{r}_1 \rho_2(\bs{r}_1, \bs{r}_1+\bs{r}),
\end{equation}
which represents the conditional probability density
of finding an electron being at $\bs{r}+\bs{R}_B$ away,
given that site 
$\bs{R}_B$ is occupied.
To compare across different phases, we normalize 
$\rho^B_2$
by the product of the total density on the $\bs{R}_B$ site $\rho_B = \int_{\bs{r}\in\mathcal{V}(\bs{R}_B)} \rho(\bs{r}) d\bs{r}$ and the mean density $\rho_0$.
As shown in Fig.~\ref{fig:dens-pair},
$\lc$ exhibits an exchange-correlation (xc) hole at short distance. The xc hole  is of similar shape in all three phases, but is reduced in the P/M because of the presence of opposite spins, and most pronounced in the F/CDW with stronger interaction. 
In the metallic phases, $\lc$ retains honeycomb sublattice symmetry away from the xc hole.
In the CDW phase,
$\lc$ exhibits long-range periodic structure consistent with a triangular lattice. 
This is most clearly seen from the tail of the $\lc$ line cuts in Fig.~\ref{fig:dens-pair}.
The imbalance between the BB and BA correlations decay with distance in the P/M and F/M phases,
while
the 
CDW phase has a persistent imbalance. 

\begin{figure}[ht]
\includegraphics[width=\linewidth]{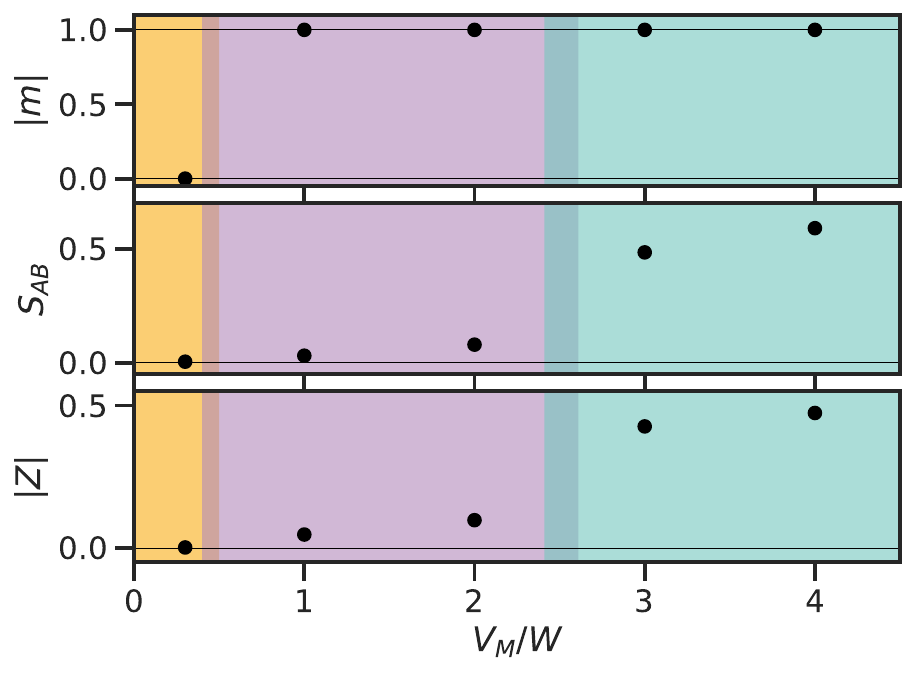}
\caption{
Quantitative measures of the magnetic and the metal-insulator transitions at $r_s=7$.
$\vert m \vert$ is the absolute magnetization per moir\'e unit cell.
$S_{AB}$ measures the degree of sublattice polarization from the two-body density matrix.
$\vert Z\vert$ is the complex polarization, which measures the degree of electron localization,
with $0$ being delocalized and $1$ being strongly localized.
The sudden changes in these quantities identify the transitions among the three phases, which demonstrate magnetic and charge orders consistent with the phase diagram.
}
\label{fig:zpol}
\end{figure}

We can define a measure for the degree of sublattice symmetry breaking using the 
imbalance in the correlation $\rho_2^B$
at larger distance:
\begin{align} \label{eq:sab}
S_{AB} =& \dfrac{\rho_2^B(\bs{R}_{B_3})-\rho_2^B(\bs{R}_{A_6})}{\rho_2^B(\bs{R}_{B_3})+\rho_2^B(\bs{R}_{A_6})}.
\end{align}
As labeled in the lower left panel of Fig.~\ref{fig:dens-pair}, the sites $\bs{R}_{B_3}$ and $\bs{R}_{A_6}$ are the third- and sixth-nearest neighbors of the reference site $\bs{R}_B$ on the $B$ and $A$ sublattices, respectively.
$S_{AB}$ close to $0$ indicates equal occupation of the two sublattices, while $S_{AB}$ close to $1$ indicates complete deoccupation of the opposite sublattice.
In order to help distinguish metallic and insulating states, we also compute 
the complex polarization~\cite{resta_electron_1999,souza_polarization_2000}
\begin{equation} \label{eq:cpol}
\vert Z \vert _N ^{\bs{g}} =
\braket{\Psi_N\vert \exp\left( {-i \sum\limits_{j=1}^{N} \bs{g}\cdot\bs{r}_j} \right)\vert\Psi_N},
\end{equation}
which is related to the degree of electron localization along the $\hat{\bs{g}}$ direction, with a value of $0$ being delocalized and $1$ being highly localized. 
In the exponent, $\bs{g}\cdot\bs{r}_j = 2\pi \sum_l m_ls_{jl}$,
the vectors $\bs{m}$ and $\bs{s}$ are fractional coordinates given by $\bs{g}=\sum_l m_l \bs{b}_l$, $\bs{r}_j=\sum_l s_{jl}\bs{a}_l$, where $\bs{a}_l$ and $\bs{b}_l$ are the direct- and reciprocal-lattice vectors of the supercell (sums run over two spatial dimensions).
As shown in Fig.~\ref{fig:zpol}, both $S_{AB}$ and $\vert Z \vert _N ^{\bs{g}}$ change abruptly across the metal-insulator transition.
The absolute magnetization, shown in the top panel of Fig.~\ref{fig:zpol}, identifies the ferromagnetic transition at a shallower moir\'e potential than the metal-insulator transition.

The appearance of large areas of ferromagnetic phases in the honeycomb lattice is in sharp contrast with 
the triangular case, where only
a paramagnetic metal to GWC insulator transition is seen in the moir\'e continuum Hamiltonian~\cite{yang_mit_2024}. 
In general, competing anti-ferromagnetic (AFM) and ferromagnetic (FM) ground states can be stabilized by the moir\'e potential.
The true ground state is determined by a delicate balance between kinetic, moir\'e, and interaction energies.
At a commensurate filling, magnetism induced by the moir\'e potential is driven by the exchange energy.
In this case, the FM phase typically hosts more localized electrons compared to the AFM phase, leading to higher kinetic and lower exchange energy.
In a triangular moir\'e potential~\cite{yang_mit_2024}, this kinetic energy increase overwhelms the gain in exchange energy, leading to an AFM ground state.
However, in the honeycomb limit, there are Dirac crossings at the $K$ points in the Brillouin zone when the system is fully polarized.
This allows the electrons to delocalize along the $\Gamma K$ directions, reducing the kinetic energy increase in the FM phase.
Further, the more delocalized electrons contribute to lower exchange energy via direct exchange with their nearest neighbors.
The F/M phase 
is robust here 
but is fragile with respect to variations in the topology.
As the moir\'e potential is tuned away from the honeycomb limit, band gaps open at the Dirac crossings, turning this phase into a band insulator.
As the potential is further tuned towards the triangular limit, direct exchange is suppressed and the lowest-energy magnetic order changes from FM to AFM. 
Detailed analysis of energetic components is provided in the supplemental materials surrounding Fig.~\ref{fig:lda-p60-p46}.

The CDW phase is stabilized by long-range Coulomb interaction, taking place much earlier (at smaller $r_s$) than in the electron gas due to  assistance from the moir\'e potential. Therefore, it can also be thought of as a GWC phase.
The electrons preferentially occupy the same 
sublattice, as shown by $\lc$ in Fig.~\ref{fig:dens-pair}.
The many-body ground state is an equal superposition of two
such states (like a ``floating crystal''~\cite{bishop_electron_1982,lewin_floating_2019}) which respects the honeycomb symmetry.
Once the F/M phase is established,
it is perhaps not too surprising to find, upon further increase of the interaction, a transition from the semimetal phase into the CDW phase.
The same phenomenon is seen in the 
spinless fermion model on a honeycomb lattice with near-neighbor interactions, where a number of many-body calculations have established a semimetal to 
CDW transition for a range of interaction strengths at half-filling~\cite{capponi_phase_2015,motruk_interaction-driven_2015}. 
This suggests that, in the moir\'e systems, the transition we have observed will likely not be affected by the presence of gates, which screen the Coloumb interaction.
It is also an indirect validation that  
Hubbard-like models (but with at least near-neighbor interactions) 
could provide a plausible way to qualitatively capture some of  the phenomena in these TMD systems.

\ssec{Conclusion and Outlook}
We have presented a comprehensive analysis of the 2DEG in the presence of a moir\'e potential with honeycomb symmetry using DMC. With increasing strength of the moir\'e potential, the system undergoes a paramagnetic-to-ferromagnetic transition where the system remains metallic, followed by a second transition to an insulating charge density wave phase. The F/M phase is characterized by a strongly anisotropic Fermi surface and 
stabilized by a combination of exchange energy gain and a lower kinetic energy penalty due to 
the more delocalized nature of electrons in the honeycomb structure. The CDW phase is characterized by a strong sublattice symmetry breaking in pair correlations, leading to a superposition of GWC states with triangular symmetry. 

The delicate interplay between band structure, moir\'e potential, and electron interactions in 2D systems leads to a rich set of interesting ordered states at low temperatures. Accurate many-body approaches are an important 
for further progress, as models become more complicated and realistic. Quantum Monte Carlo approaches offer an excellent balance between accuracy and scalability, allowing 
continuum models to be treated without 
further approximations while reaching sufficiently large system sizes that enable more reliable extrapolation to 
the thermodynamic limit.
A variety of related systems are accessible either directly with our approach or by introducing reasonably straightforward variations.
Given the rapid pace at which experiments are progressing 
we hope this work will help open up more
broad applications of accurate many-body computations in this area. 

\ssec{Acknowledgment}
The Flatiron Institute is a division of the Simons Foundation.
We thank Daniele Guerci, Giorgio Sangiovanni, Valentin Crepel, Yang Zhang, and Yixiao Chen for useful discussions.

\bibliography{ref}

\clearpage
\onecolumngrid
\renewcommand{\thefigure}{S\arabic{figure}}
\setcounter{figure}{0}
\section{Ferromagnetic semimetal and charge density wave phases of interacting electrons in a honeycomb moir\'e potential : Supplemental Materials}

\subsection{Momentum Distribution}

\begin{figure}[ht]
\begin{minipage}{0.48\linewidth}
\includegraphics[width=\linewidth]{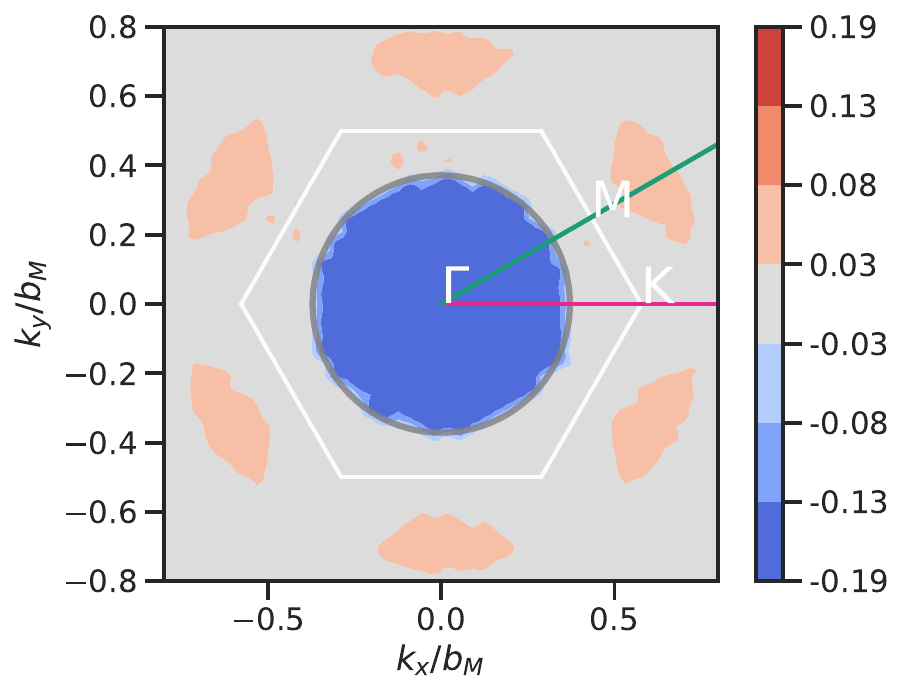}
\end{minipage}
\begin{minipage}{0.48\linewidth}
\includegraphics[width=\linewidth]{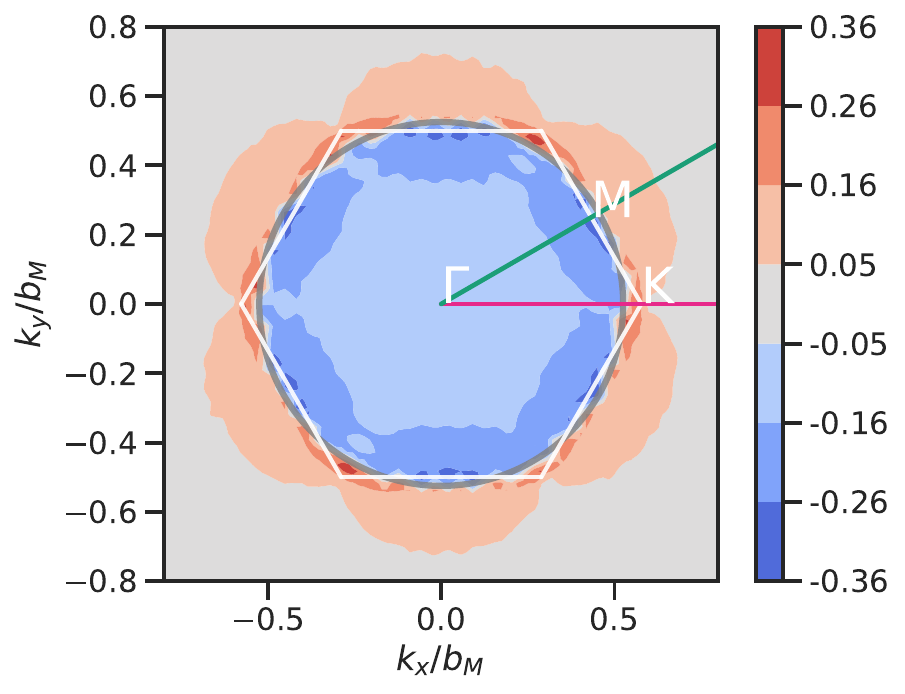}
\end{minipage}
\begin{minipage}{0.48\linewidth}
\includegraphics[width=\linewidth]{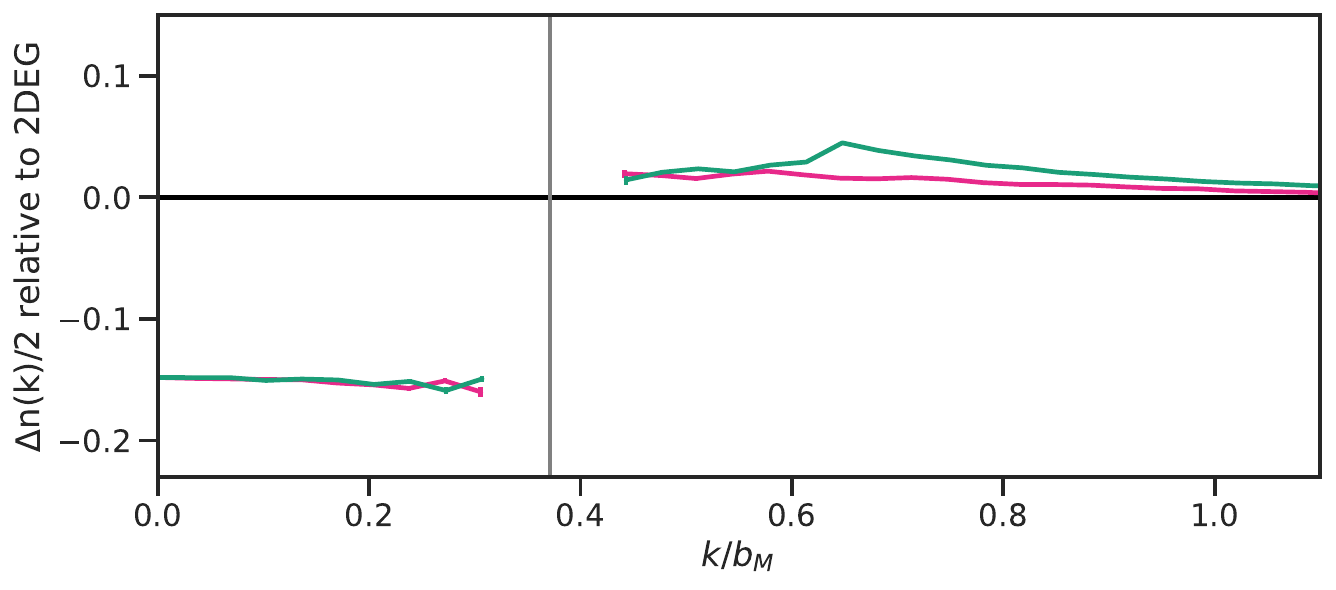}
\end{minipage}
\begin{minipage}{0.48\linewidth}
\includegraphics[width=\linewidth]{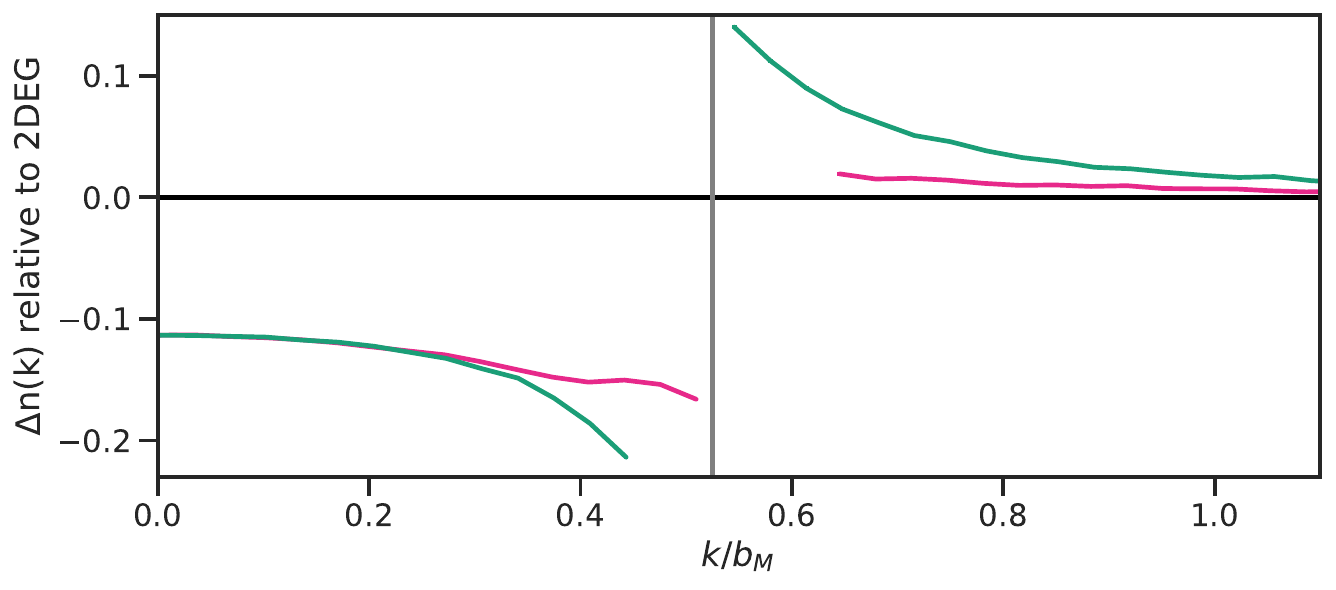}
\end{minipage}
\begin{minipage}{0.48\linewidth}
\includegraphics[width=\linewidth]{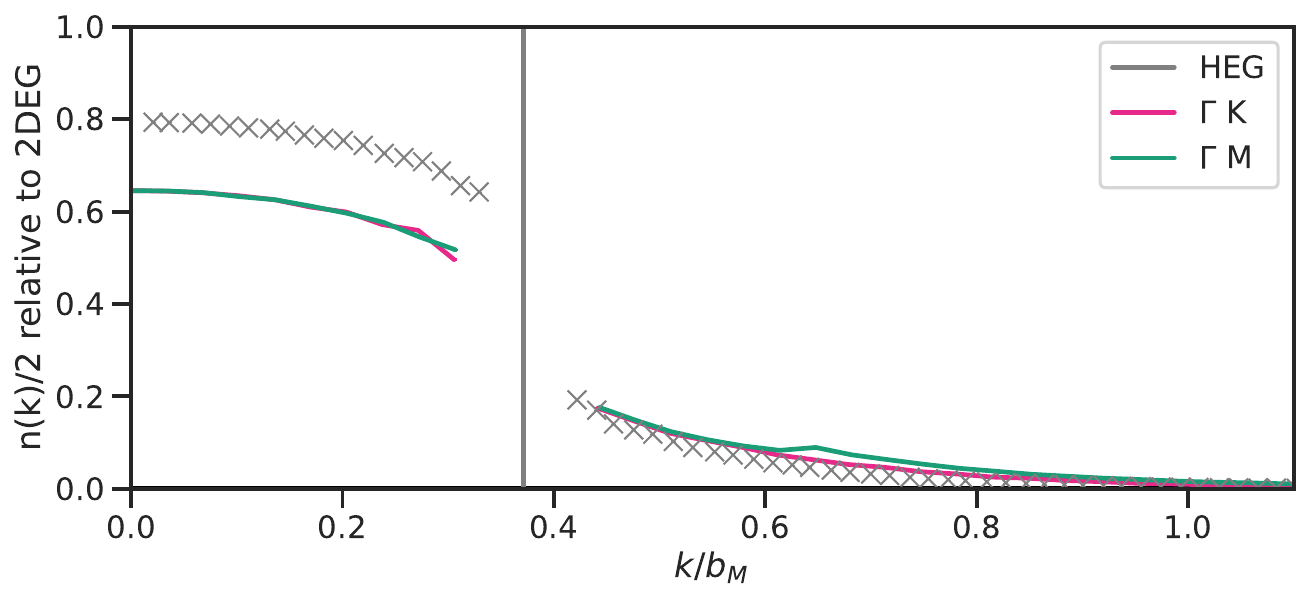}
\end{minipage}
\begin{minipage}{0.48\linewidth}
\includegraphics[width=\linewidth]{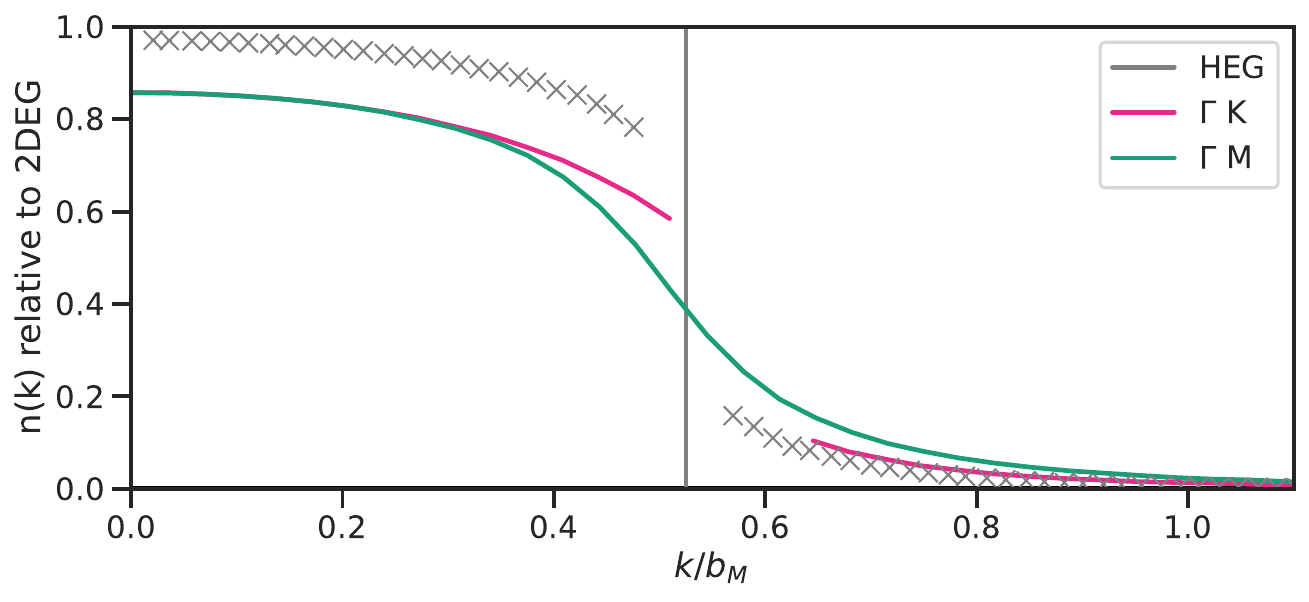}
\end{minipage}
\caption{Redistribution of the electronic momentum density of the two-dimensional electron gas due to the honeycomb moir\'e potential in the (left) P/M and (right) F/M phases both stablized at $\lambda=1.5$.
Each simulation contains $N=144$ electrons with $2\times 2$ twists in the canonical ensemble, which introduces some noise close to the Fermi surface.
In the line cuts, points too close to the Fermi surface are excluded.
}
\label{fig:dmc-p60-nk-heg}
\end{figure}

Figure~\ref{fig:dmc-p60-nk-heg} shows how the moir\'e potential redistributes the momentum density 
in the P/M and F/M phases, 
relative to that of the 2DEG.
In the P/M phase, the moir\'e potential scatters low-momentum states from within the Fermi surface to the high-momentum tail in an isotropic manner, except when close to secondary Fermi surfaces.
In contrast, the redistribution of momentum density in the F/M phase near the 2DEG Fermi surface is highly anisotropic.
Along the $\Gamma K$ direction, a similar amount of momentum density is moved across the Fermi surface as in the P/M phase.
However, along the $\Gamma M$ direction, significantly more momentum density is transferred to make $n(k)$ smooth.

\subsection{Determination of Transition Boundaries}

\begin{figure}[ht]
\begin{minipage}{0.48\textwidth}
\includegraphics[width=\linewidth]{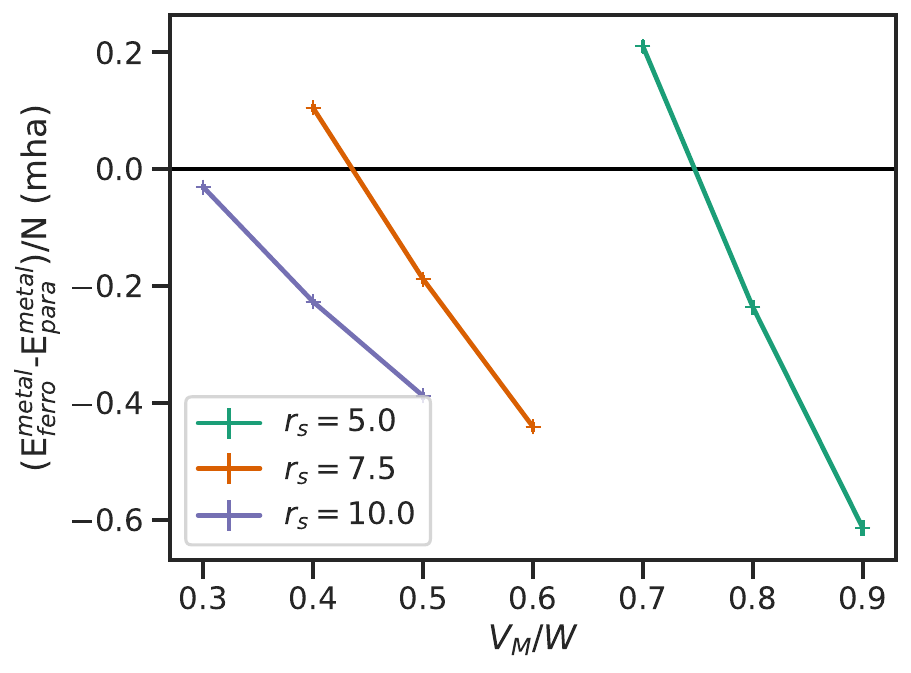}
(a) paramagnetic (P/M) to ferromagnetic (F/M)
\end{minipage}
\begin{minipage}{0.48\textwidth}
\includegraphics[width=\linewidth]{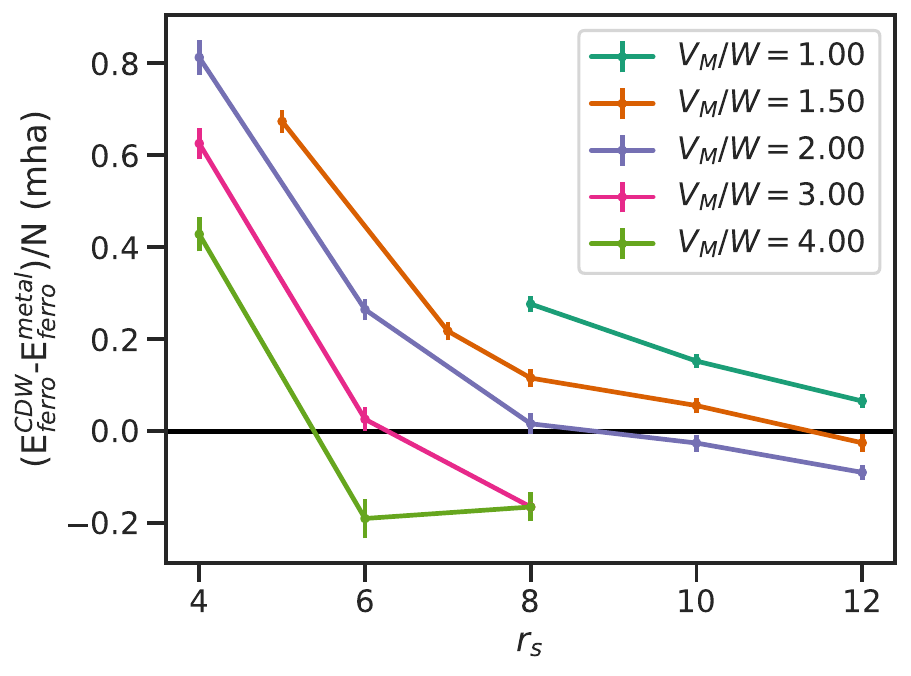}
(b) metal (F/M) to insulator (F/CDW)
\end{minipage}
\caption{Transitions determined by total energy difference.}
\label{fig:rsmu-detot}
\end{figure}

The ferromagnetic transition from the P/M to the F/M phase and the metal-insulator transition from the F/M to the F/CWD phase are obtained by comparing total energies of candidate states.
We perform meta-stable simulations of the three phases at a range of $r_s$ and $\lambda=V_M/W$ conditions and find the ground state using the variational property of DMC.
Non-interacting orbitals are used in the determinant part of the wavefunction for the P/M and F/M phases, whereas Hartree-Fock (HF) orbitals are used for the F/CDW phase.
The ferromagnetic transitions are obtained from Fig.~\ref{fig:rsmu-detot}(a), whereas the metal-insulator transitions are extracted from Fig.~\ref{fig:rsmu-detot}(b).
Twist-averaged DMC total energies are tabulated in Table~\ref{tab:rsmu}.
All calculations used to determine the transitions have been performed in simulation cells containing $N=144$ electrons with $1024$ walkers.
The total energy is averaged over a shifted uniform grid of twists ($2\times 2$) in the canonical ensemble.

\begin{table}[h]
\caption{DMC total energies used in Fig.~\ref{fig:rsmu-detot}.}
\label{tab:rsmu}
\hfill
\begin{tabular}{lrrl}
\toprule
phase & $r_s$ & $\lambda$ & $E_{tot}$ \\
\midrule
F/M & 5.0 & 0.7 & -0.165712(2) \\
P/M & 5.0 & 0.7 & -0.16592(1) \\
F/M & 5.0 & 0.8 & -0.170379(2) \\
P/M & 5.0 & 0.8 & -0.17014(1) \\
F/M & 5.0 & 0.9 & -0.175261(2) \\
P/M & 5.0 & 0.9 & -0.17465(1) \\
F/M & 7.5 & 0.4 & -0.111334(1) \\
P/M & 7.5 & 0.4 & -0.111438(6) \\
F/M & 7.5 & 0.5 & -0.113097(2) \\
P/M & 7.5 & 0.5 & -0.112909(7) \\
F/M & 7.5 & 0.6 & -0.114997(2) \\
P/M & 7.5 & 0.6 & -0.114555(9) \\
F/M & 10.0 & 0.3 & -0.086346(1) \\
P/M & 10.0 & 0.3 & -0.086315(5) \\
F/M & 10.0 & 0.4 & -0.087283(1) \\
P/M & 10.0 & 0.4 & -0.087056(5) \\
F/M & 10.0 & 0.5 & -0.088309(2) \\
P/M & 10.0 & 0.5 & -0.087920(7) \\
\bottomrule
\end{tabular}

\hfill
\begin{tabular}{lrrl}
\toprule
phase & $r_s$ & $\lambda$ & $E_{tot}$ \\
\midrule
F/CDW & 8.0 & 1.0 & -0.11614(2) \\
F/M & 8.0 & 1.0 & -0.116412(5) \\
F/CDW & 10.0 & 1.0 & -0.09429(1) \\
F/M & 10.0 & 1.0 & -0.094444(9) \\
F/CDW & 12.0 & 1.0 & -0.07948(1) \\
F/M & 12.0 & 1.0 & -0.07954(1) \\
F/CDW & 5.0 & 1.5 & -0.20753(3) \\
F/M & 5.0 & 1.5 & -0.208137(4) \\
F/CDW & 8.0 & 1.5 & -0.12742(2) \\
F/M & 8.0 & 1.5 & -0.12753(1) \\
F/CDW & 10.0 & 1.5 & -0.10165(1) \\
F/M & 10.0 & 1.5 & -0.10171(1) \\
F/CDW & 12.0 & 1.5 & -0.084688(8) \\
F/M & 12.0 & 1.5 & -0.08466(2) \\
F/CDW & 4.0 & 2.0 & -0.31168(4) \\
F/M & 4.0 & 2.0 & -0.312489(5) \\
F/CDW & 6.0 & 2.0 & -0.19298(2) \\
F/M & 6.0 & 2.0 & -0.193249(9) \\
\bottomrule
\end{tabular}

\hfill
\begin{tabular}{lrrl}
\toprule
phase & $r_s$ & $\lambda$ & $E_{tot}$ \\
\midrule
F/CDW & 8.0 & 2.0 & -0.13979(2) \\
F/M & 8.0 & 2.0 & -0.13980(2) \\
F/CDW & 10.0 & 2.0 & -0.109696(9) \\
F/M & 10.0 & 2.0 & -0.10967(2) \\
F/CDW & 12.0 & 2.0 & -0.090353(8) \\
F/M & 12.0 & 2.0 & -0.09026(1) \\
F/CDW & 4.0 & 3.0 & -0.41749(3) \\
F/M & 4.0 & 3.0 & -0.418111(8) \\
F/CDW & 6.0 & 3.0 & -0.24056(2) \\
F/M & 6.0 & 3.0 & -0.24059(2) \\
F/CDW & 8.0 & 3.0 & -0.16689(1) \\
F/M & 8.0 & 3.0 & -0.16672(2) \\
F/CDW & 4.0 & 4.0 & -0.53228(3) \\
F/M & 4.0 & 4.0 & -0.53271(1) \\
F/CDW & 6.0 & 4.0 & -0.29210(2) \\
F/M & 6.0 & 4.0 & -0.29191(4) \\
F/CDW & 8.0 & 4.0 & -0.196124(7) \\
F/M & 8.0 & 4.0 & -0.19596(3) \\
\bottomrule
\end{tabular}

\hfill
\end{table}

\subsection{Superposition of Wavefunctions}

\begin{figure}[ht]
\begin{minipage}{.6\textwidth}
\includegraphics[width=\linewidth]{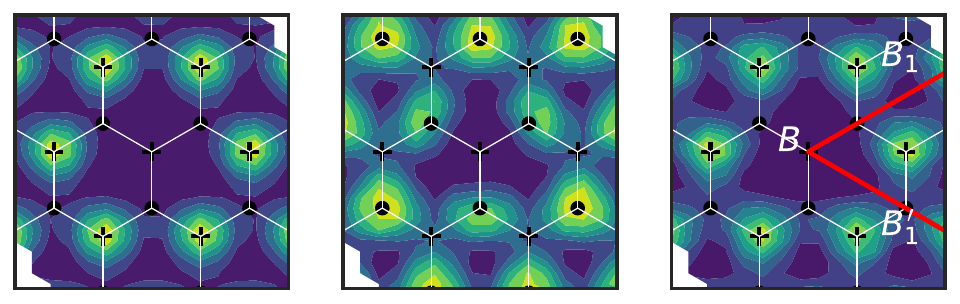}
\end{minipage}
\begin{minipage}{.6\textwidth}
\includegraphics[width=\linewidth]{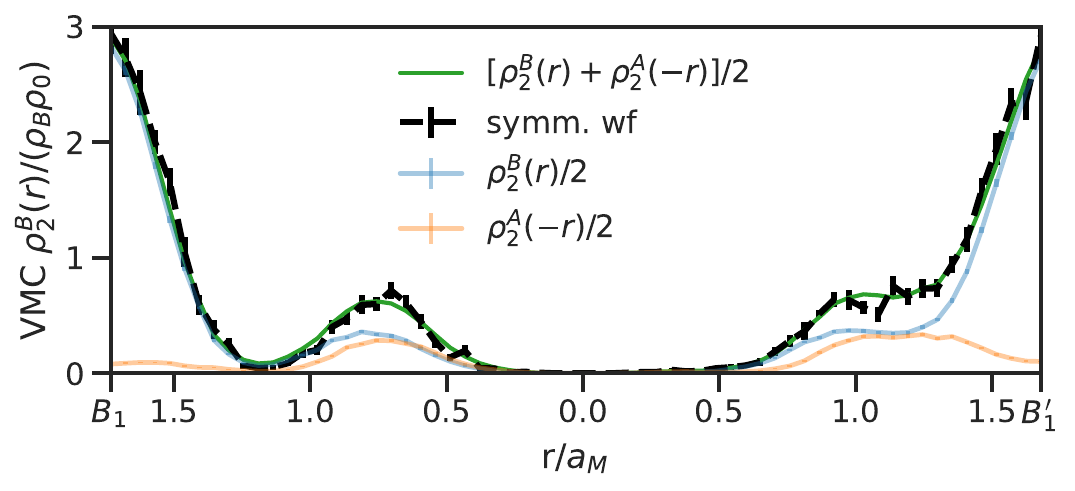}
\end{minipage}
\caption{Symmetrization of pair correlation matches wavefunction superposition at $r_s=7$ and $V_M/W=3$.
Test calculations were performed with $N=16$ electron in a simulation cell with periodic boundary conditions.}
\label{fig:symm-pair}
\end{figure}

The ground state of the F/CDW phase retains honeycomb sublattice symmetry of the Hamiltonian.
However, the HF determinant used in our trial wavefunction breaks this symmetry.
To restore it for all properties, we use the equal superposition of two HF solutions, one occupying the A sublattice while the other occupying B, in the determinant part of the trial wavefunction
\begin{align}
\Psi_T = \left[ \Psi_{HF}^A + \Psi_{HF}^B \right] \exp(-U),
\end{align}
where $\exp(-U)$ is the Jastrow.
Alternatively, we can restore sublattice symmetry of the density and pair correlation by averaging over elements of the inversion group with generator $\mathcal{I}$, where $\mathcal{I} \rho(\bs{r})=\rho(-\bs{r})$ and $\mathcal{I} \rho_2(\bs{r}_1, \bs{r}_2) = \rho_2(-\bs{r}_1, -\bs{r}_2)$. Thus, $\mathcal{I}\rho_2^B(\bs{r}) = \rho_2^A(-\bs{r})$.
The charge density and pair correlation of eq.~(\ref{eq:rho2a}) can be symmetrized as $\rho(\bs{r}) \rightarrow [\rho(\bs{r})+\mathcal{I}\rho(\bs{r})]/2$ and $\rho_2^B(\bs{r}) \rightarrow [\rho_2^B(\bs{r})+\mathcal{I}\rho_2^B(\bs{r})]/2$, respectively.
As shown in Fig.~\ref{fig:symm-pair}, the two approaches produced identical results at $r_s=7$ and $V_M/W=3$.
We directly symmetrized the properties shown in Fig.~\ref{fig:dens-pair}.

\subsection{Comparison with Density Functional Theory}

Once the phase diagram Fig.~\ref{fig:phase} is established by the quantum Monte Carlo calculations, we also attempted to reproduce it using density functional theory.
Using LDA, we find the paramagnetic metal to the ferromagnetic semimetal transition, but the band crossings at the $K$ points remain at $V_M/W$ as large as $8$.
After mixing $50$\% of exact exchange, we find a direct transition from the paramagnetic metal to the ferromagnetic insulator.
We were not able to obtain both the ferromagnetic semimetal and the insulator at any value of exact exchange fraction.

\begin{figure}[ht]
\begin{minipage}{0.32\linewidth}
\includegraphics[width=\linewidth]{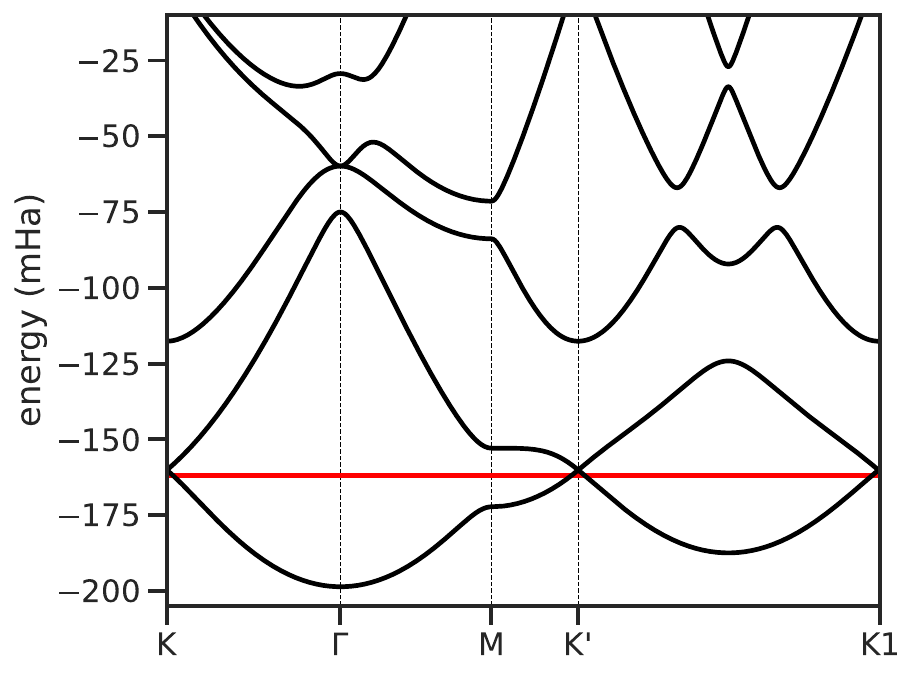}
\end{minipage}
\begin{minipage}{0.32\linewidth}
\includegraphics[width=\linewidth]{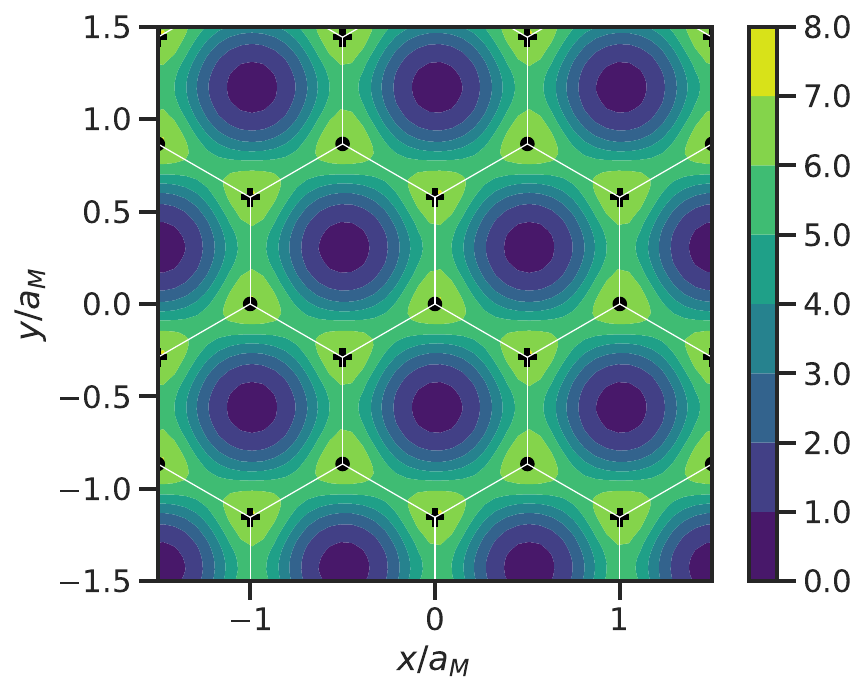}
\end{minipage}
\begin{minipage}{0.32\linewidth}
\includegraphics[width=\linewidth]{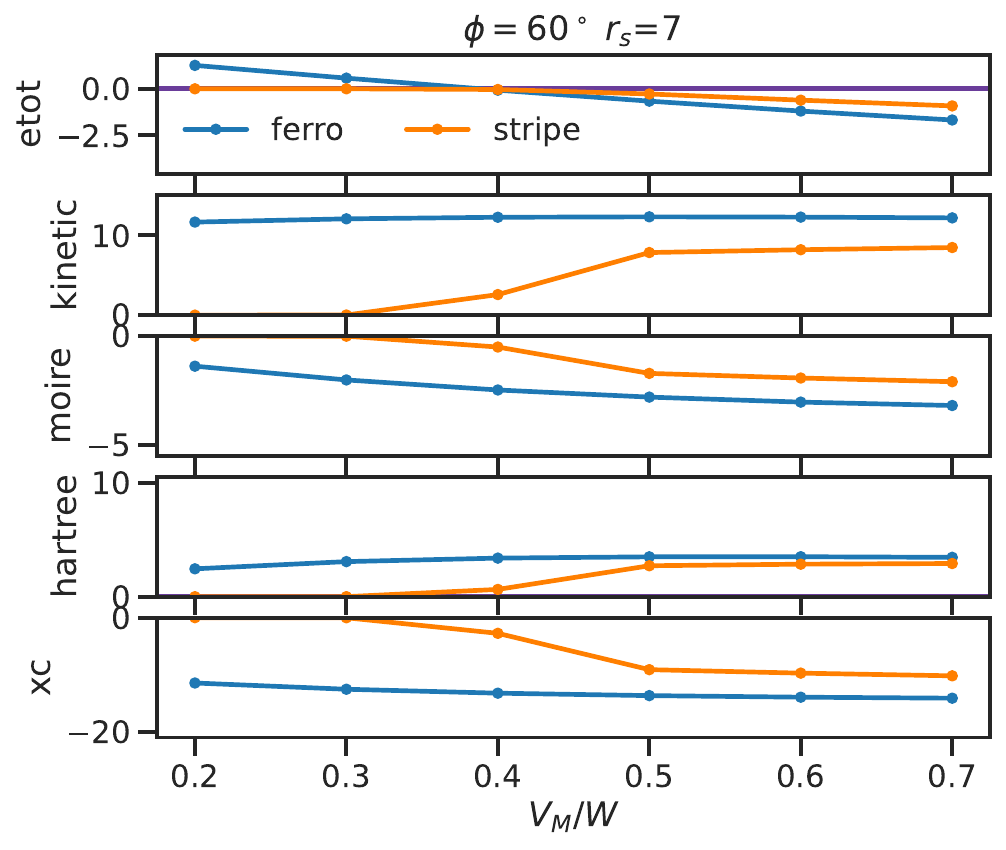}
\end{minipage}

\begin{minipage}{0.32\linewidth}
\includegraphics[width=\linewidth]{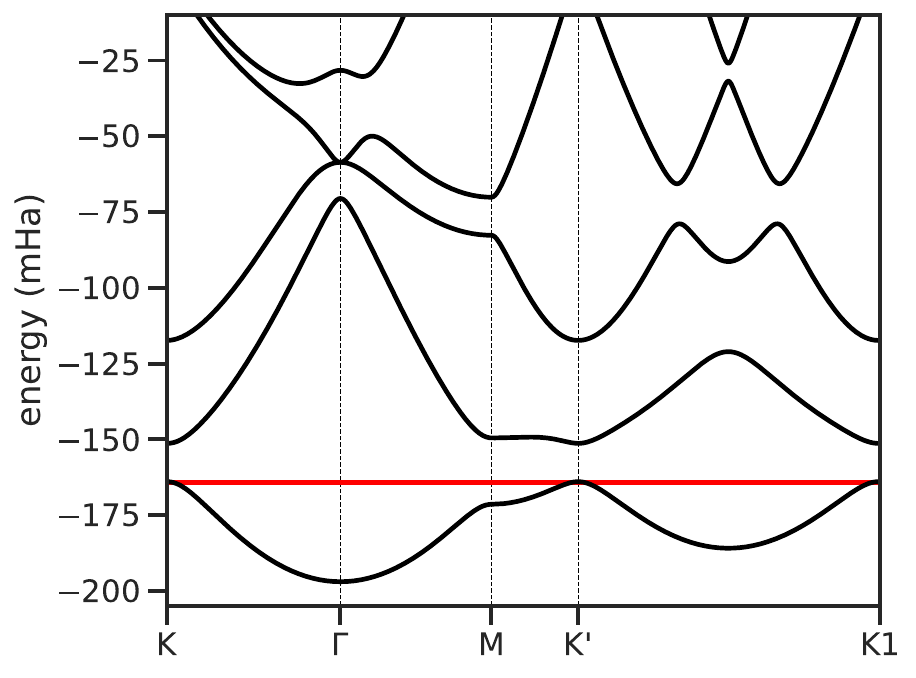}
\end{minipage}
\begin{minipage}{0.32\linewidth}
\includegraphics[width=\linewidth]{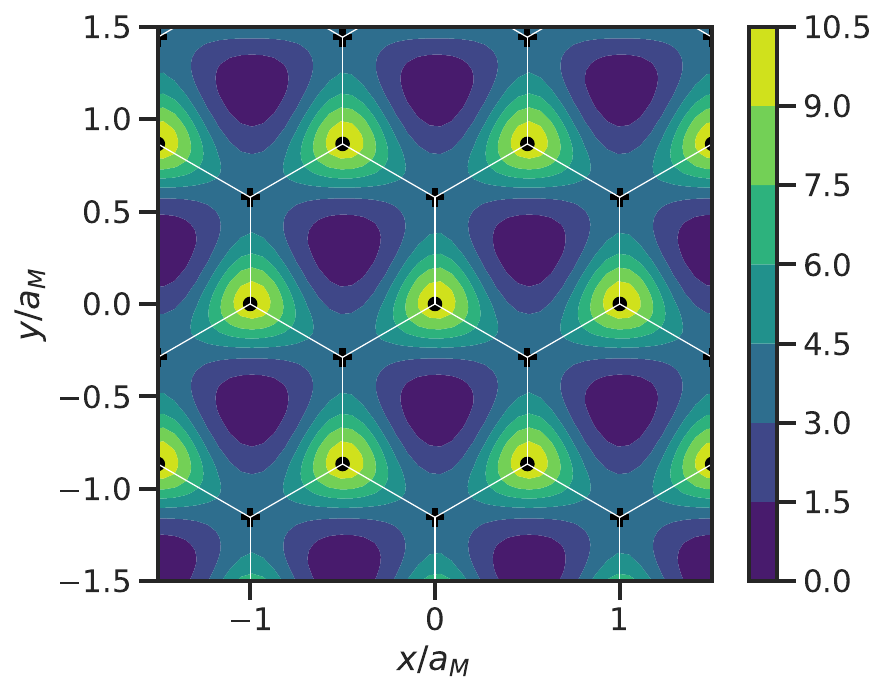}
\end{minipage}
\begin{minipage}{0.32\linewidth}
\includegraphics[width=\linewidth]{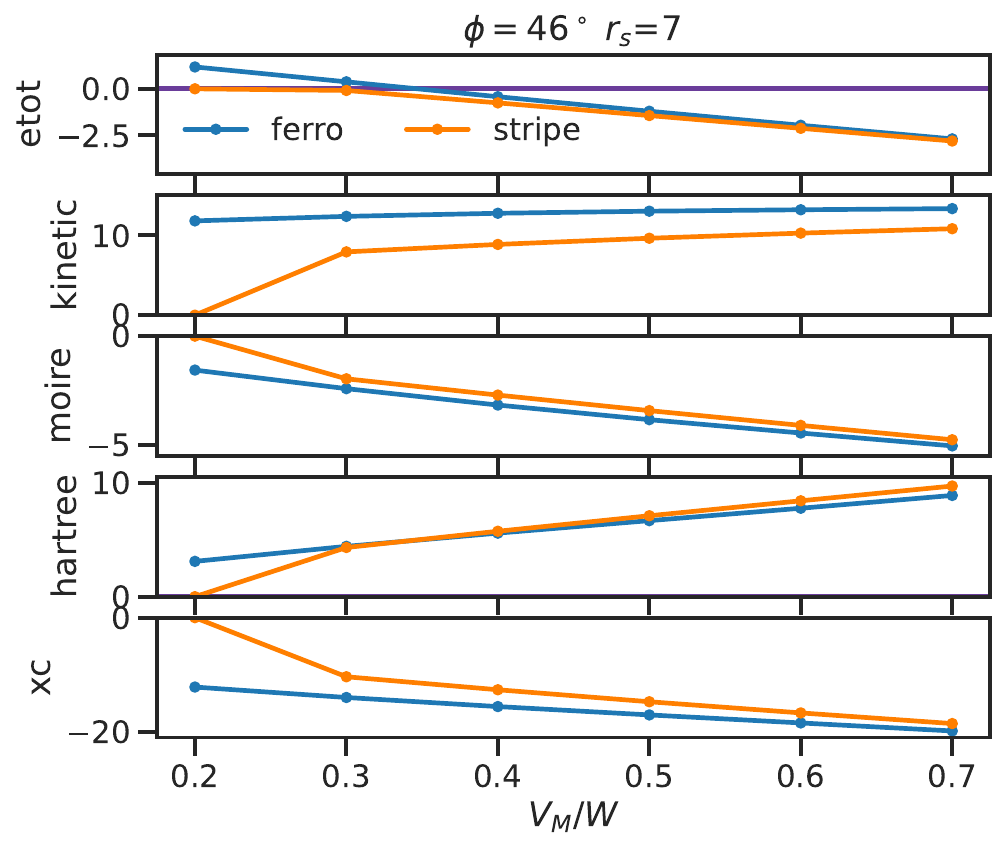}
\end{minipage}

\caption{LDA charge density (left panel) of the ferromagnetic state at $V_M/W=0.5$ and the corresponding band structure (middle panel) in two moir\'e potentials, one in the honeycomb limit ($\phi=60^\circ$ shown in the top row), while the other tuned towards the triangular limit, which is at $\phi=0^\circ$.
Black dots and crosses label the honeycomb A and B sublattices, respectively.
Energy components (right column) of the FM phase is compared against those in
a collinear phase with anti-ferromagnetic spin stripes, which is used as a proxy for the AFM phase.
}
\label{fig:lda-p60-p46}
\end{figure}

Since the P/M to F/M transition can be captured by DFT calculations using the 2D LDA functional~\cite{attaccalite_correlation_2002},
we use it to perform 
detailed analysis of the band structure and various energy components to help 
identify the driving mechanism of this transition.
We tune the moir\'e potential from the honeycomb limit ($\phi=60^\circ$), where we find a ferromagnetic (FM) phase, towards the triangular limit ($\phi=0^\circ$), where an anti-ferromagnetic (AFM) phase is expected~\cite{yang_mit_2024}.
The charge density, LDA band structure and energy components are shown in Fig.~\ref{fig:lda-p60-p46}.
In a honeycomb moir\'e potential, the LDA band structure always has a band crossing at the $K$ points of the Brillouin zone and a charge density with honeycomb symmetry.
However, at $\phi=46^\circ$, gaps open at the $K$ points and the charge density has only triangular symmetry.
Therefore, the honeycomb moir\'e potential allows the electronic wavefunction to delocalize more uniformly across the two local minima, which enhances the direct exchange interaction.
According to the energy components in Fig.~\ref{fig:lda-p60-p46}, as the system exits the P/M phase around $V_M/W=0.5$, the FM and AFM phases have nearly identical Hartree energy.
The FM phase has lower moir\'e and exchange-correlation (xc) components than the AFM phase and higher kinetic energy.
When the moir\'e potential has honeycomb symmetry, the kinetic energy of the FM phase is insensitive to its strength ($V_M/W$).
Further, the moir\'e and xc components are significantly lower than the AFM phase compared to the $\phi=46^\circ$ case.
This together with the changes in the charge density suggest %
that honeycomb moir\'e potential favors FM by strengthening the exchange interaction and better utilizing the moir\'e potential at %
reduced cost to the kinetic energy.

\end{document}